# Selective control of localised vs. delocalised carriers in anatase TiO$_2$ through reaction with O$_2$

Chiara Bigi[1,2], Zhenkun Tang[3,4], Gian Marco Pierantozzi[1], Pasquale Orgiani[1,5], Pranab Kumar Das[1], Jun Fujii[1], Ivana Vobornik[1], Tommaso Pincelli[1], Alessandro Troglia[1,2], Tien-Lin Lee[6], Regina Ciancio[1], Goran Dražic[7], Alberto Verdini[1], Anna Regoutz[8], Phil D. C. King[9], Deepnarayan Biswas[9], Giorgio Rossi[1,2], Giancarlo Panaccione[1*] and Annabella Selloni[3*]

[1] *Istituto Officina dei Materiali (IOM)-CNR, Laboratorio TASC, in Area Science Park, S.S.14, Km 163.5, I-34149 Trieste, Italy*

[2] *Dipartimento di Fisica, Università di Milano, Via Celoria 16, I-20133 Milano – Italy*

[3] *Department of Chemistry, Princeton University, Princeton, New Jersey 08544, USA*

[4] *Department of Physics, Hengyang Normal University, 16 Heng-Hua Road, Zhu-Hui District, Hengyang, P. R. China, 421008*

[5] *CNR-SPIN, UOS Salerno, 84084 Fisciano, Italy*

[6] *Diamond Light Source, Harwell Science and Innovation Campus, Didcot OX11 0DE, United Kingdom*

[7] *Department for Materials Chemistry, National Institute of Chemistry, Hajdrihova 19, SI- 1001 Ljubljana, Slovenia*

[8] *Department of Materials, Imperial College London, South Kensington, London SW7 2AZ, United Kingdom*

[9] *SUPA, School of Physics and Astronomy, University of St. Andrews, St. Andrews KY16 9SS, United Kingdom*

*Corresponding authors: aselloni@Princeton.edu (A.S.), giancarlo.panaccione@elettra.eu (G.P)

SUMMARY

Anatase titanium dioxide ($TiO_2$) is one of the most widely used metal oxides for applications ranging from photo-catalysis for water splitting and environmental remediation to self-cleaning windows and biomedical devices. In stoichiometric form, $TiO_2$ has a large band gap, ~ 3.2 eV, and is photo-active only under UV light illumination, which accounts for just 5% of the solar spectrum. However, $TiO_2$ can be easily reduced through the creation of oxygen vacancies. These give rise to localized defect states within the band-gap (in-gap (IG)), extending $TiO_2$'s photo-response to the visible light region. In addition, as found in other oxide systems, oxygen vacancies in anatase induce also highly delocalized (conductive) two-dimensional electron gas (2DEG) states in proximity of the surface. While the influence of oxygen vacancies on $TiO_2$'s photo-activity is well known, it is still unclear whether the effects of IG and 2DEG states are similar or not and which is more or less favorable. For this reason, one would ideally like to control and selectively activate one of the two states (i.e. IG *or* 2DEG). Such a control has not yet been realized in any of the wide band-gap oxides.

To achieve this objective, we have investigated the oxygen vacancy induced IG and 2DEG states on the anatase $TiO_2$ (001) surface using a combined experimental and computational approach. Our first principles calculations show that the IG and 2DEG states in $TiO_2$ originate from different vacancy sites: the former from sites localized directly at the surface, and therefore highly reactive to oxygen; the latter are instead subsurface and much more robust to oxygen. Based on these results, we devised a way to selectively control the localized IG states by reaction with molecular oxygen, as confirmed by operando experiments under oxygen dosing. The fact that the 2DEG states are located at well-defined sub-surface layers opens up the possibility of their direct control by material atomic-engineering.

By unveiling new and relevant aspects of the surface chemistry of TiO2, our findings provide a pathway for tailoring the performance of future devices.

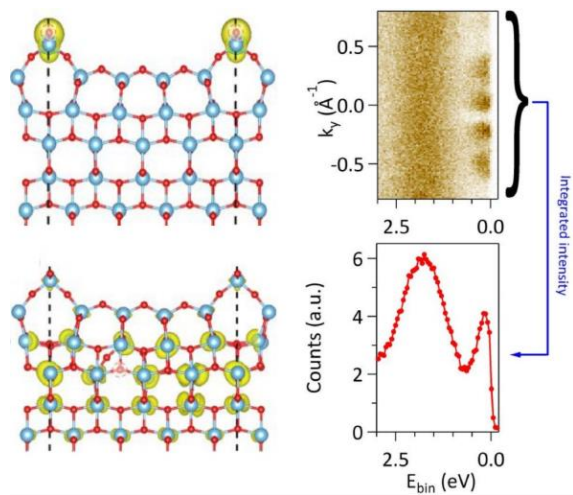

ABSTRACT

Two-dimensional (2D) metallic states induced by oxygen vacancies at oxide surfaces and interfaces provide new opportunities for the development of advanced applications, but the ability to control the behavior of these states is still limited. We used Angle Resolved Photoelectron Spectroscopy combined with density functional theory to study the reactivity of states induced by the oxygen vacancies at the (001)–(1×4) surface of anatase $TiO_2$, where both 2D metallic and deeper lying in-gap states (IGs) are observed. Remarkably, the two states exhibit very different evolution when the surface is exposed to molecular $O_2$: while IGs are almost completely quenched, the metallic states are only weakly affected. The energy scale analysis for the vacancy migration and recombination resulting from the DFT calculations confirms indeed that only the IGs originate from and remain localized at the surface, whereas the metallic states originate from subsurface vacancies, whose migration and recombination at the surface is energetically less favorable rendering them therefore insensitive to oxygen dosing.

ARTICLE TEXT

I.  INTRODUCTION

Many functional properties of anatase $TiO_2$ of relevance, e.g., in photocatalysis, solar cells and sensors – are critically affected by the presence of excess electrons induced by intrinsic defects, dopants or photoexcitation[1-7]. Understanding and controlling the behaviour of excess electrons is thus essential for improving its performance in existing applications and for developing new applications as well. In particular, important changes in the electronic structure have been induced by the chemical doping arising from oxygen vacancies ($V_{OS}$), such as the creation of in-gap defect states, the formation of depletion regions, and band bending.[3, 8] Another noteworthy feature connected with $V_{OS}$ is the formation, under photoirradiation, of electronic states with metallic $d$-character (Ti $3d$) at the anatase (101) and (001) surfaces, generally termed two-dimensional electron gas (2DEG) states[9-13]. First observed at the $LaAlO_3/SrTiO_3$ interface[14], 2DEGs have been reported both in transition metal oxides (TMOs) parent compounds (e.g. bare surfaces of $SrTiO_3$, $KTiO_3$) and in engineered heterostructures[15, 16]. Although the photoirradiation process is the common denominator of 2DEGs in the former materials, including anatase $TiO_2$, a number of important aspects are still controversial and need clarification, such as i) the depth distribution of the oxygen vacancies acting as electron donors, and ii) the behaviour of both 2DEG and localised in-gap (IG) defect states under reducing vs. oxidising conditions, notably to what extent and in what conditions the excess of $V_O$s created by photoirradiation can be controlled [9,10].

Here we combine ultraviolet (UV) and X-ray based electron spectroscopies and first principles calculations to clarify the role, the formation mechanism and the possible control of defect states formed at the (001) surface of anatase $TiO_2$[17-19]. In-situ UHV growth of high quality epitaxial thin films obtained by Pulsed Laser Deposition (PLD) allowed us to identify the types of defect states that appear under photoirradiation and can be tailored also by controlled post-growth treatment (UHV annealing). Angle Resolved PhotoElectron Spectroscopy (ARPES) confirms the existence of both localised and delocalised electronic states, while Resonant-PES in soft X-ray range identifies their $Ti^{4+}$ or $Ti^{3+}$ character, respectivley. Monitoring the spectral changes while dosing the surface with molecular $O_2$ reveals that the 2DEG delocalized features are robust against oxygen exposure, whilst the localised IG states get suppressed. Comparison with Density Functional Theory (DFT) calculations provides evidence of a distinct depth-dependence of defect states. The 2DEG originates from subsurface $V_O$s and resides in sub-surface layers due to the attractive potential resulting from these $V_O$s. In contrast, the deeper lying IG states that are suppressed by $O_2$ are originating from the surface $V_O$s. Our results also provide a consistent explanation of previous contrasting findings and suggest possible strategies for controlling the carriers' concentration and transport at the surface of anatase-based materials.

## II. EXPERIMENTAL RESULTS 1 – ANATASE $TiO_2$ GROWTH, STRUCTURE AND ANGLE-RESOLVED PHOTOEMISSION (ARPES)

### A. EXPERIMENTAL METHODS

#### A.1 GROWTH

Anatase $TiO_2$ thin films were grown by Pulsed Laser Deposition (PLD) at a dedicated chamber located at the APE-IOM laboratory (NFFA facility, Trieste, Italy). Rutile $TiO_2$ single-crystal was ablated using a KrF excimer pulsed laser source kept at about 2 J/cm2 energy density, with a typical laser repetition rate of 3 Hz. The substrate was kept at 700°C growth temperature, while oxygen background pressure was set to 10−4 mbar. Annealed samples have been kept at the growth temperature for 10 minutes in UHV (PLD chamber base pressure is the range of 10-7mbar). Anatase $TiO_2$ thin films were grown on (001) $LaAlO_3$ and (001) Nb-doped $SrTiO_3$ substrates. Epitaxial strain-less condition was verified for anatase $TiO_2$ thin films grown on $LaAlO_3$ substrates.

A.2 TRANSMISSION AND SCANNING TRANSMISSION ELECTRON MICROSCOPY

Cs probe-corrected Jeol ARM 200 CF scanning transmission electron microscope with cold-FEG electron source, operated at 200 kV was used for high-resolution imaging of the samples. Electron Energy Loss Spectroscopy (EELS) was performed using Gatan dual-EELS Quantum ER system and elemental chemical analyses were performed with Centurio Jeol Energy Dispersive X-ray Spectroscopy (EDXS) system with 100 mm2 SDD detector. Cross-sectional samples in the [010] zone axis suitable for TEM/STEM analyses have been obtained by a conventional polishing technique followed by dimpling and ion milling.

A.3 ULTRAVIOLET ARPES, SOFT-X RAY ARPES, RESPES

The as-grown samples were directly transferred in-situ to the Angle-Resolved Photoemission (ARPES) end-station installed on the Low-Energy branch of APE beamline (APE-LE) at Elettra synchrotron (Trieste, Italy). Such a chamber is equipped with a Scienta DA30 hemispherical electron energy and momentum analyser (30° angular acceptance), which allows to map the electronic bands over the extended areas of the Brillouin zone without rotating the smaple. ARPES experiments were performed at a base pressure $< 10^{-10}$ mbar and with the samples kept at liquid Nitrogen. Photon energy of 46 eV was used with the light incidence angle of 45°. All the light polarisation available at the beamline have been exploited (linear vertical, linear horizontal, circular right and circular left). When not otherwise specified, the overall energy resolution was set to ~40 meV, and the angular resolution was set to 0.2° (corresponding to ~0.01 Å$^{-1}$ at 46 eV photon energy).

Soft x-ray ARPES, Resonant Photoemission (RESPES) and oxygen dosing were performed at I09 beamline at Diamond light source (Didcot, UK). Sample temperature was 90 K. In order to reduce the effects of higher order components coming from the beamline optics, the monochromator has been tuned to obtain the best compromise between flux, resolution and higher order rejection. Furthermore, the residual second-order contribution was subtracted in all spectra. The energy position of the Fermi energy ($E_F$) and the energy resolution have been estimated by measuring the Fermi edge of poly-Au foil in thermal and electric contacts with the sample. The overall energy resolution (analyser + beamline) was kept below 250 meV for the entire photon energy range. Molecular oxygen was injected through a metallic capillary placed close to the sample surface. The amount of oxygen has been monitored by means of a Residual Gas Analyser available in the experimental chamber. Base pressure in the experimental chamber was $1 \times 10^{-10}$ mbar, up to a maximum $O_2$ partial pressure of $4 \times 10^{-9}$ mbar.

## B. STRUCTURAL AND ARPES RESULTS

Epitaxial strainless anatase $TiO_2$ thin films were grown by PLD on $LaAlO_3$ substrates. Consistent results were also obtained for films grown on Nb-doped $SrTiO_3$, indicating that the observed properties are intrinsic to the $TiO_2$ anatase phase (see Figure S1). Details of the growth protocol and characterization results[20] are given in Supplemental material. The results of our cross-sectional high-resolution Transmission Electron Microscopy (TEM) and high-angle annular dark-field scanning TEM (HAADF-STEM) measurements are shown in Figure 1. In panel a), a representative high-resolution Z-contrast image shows an atomically sharp interface region. The crystal quality of the films extends up to the surface, as confirmed by the Low Energy Electron Diffraction (LEED) (1×4)-(4×1) surface reconstruction pattern in panel b)[9, 11, 20].

ARPES measurements were performed along the $\bar{\Gamma} - \bar{X}$ direction of the surface projected Brillouin zone (Figure 1 c). The surface structural reconstruction is reflected in the Fermi surface measured in the first Brillouin zone, shown in panel d): The bright circle centred at the $\bar{\Gamma}$ point corresponds to a 2DEG, characterised by a parabolic dispersion (panels e) and f)) and a $d_{xy}$ orbital character arising from Ti $3d$ states typical in TMOs[9, 11, 15, 21]. In addition, several replicas occur along both $k_x$ and $k_y$ directions, arising from the periodic lateral perturbation induced by the surface (1×4)-(4×1) reconstruction[11, 19]. Panel e) and f) compare the E vs. k dispersion of the 2DEG for UHV annealed and as-grown samples respectively. Both the as-grown and reduced samples were transferred in-situ in UHV (i.e. without exposure to air) for the ARPES experiments. Two dispersive parabolic-like states are evident at $\bar{\Gamma}$ in Figure 1e: an outer parabola and a second (fainted) quantized sub-band related to the confinement potential at the surface[11, 12]. The bottom of the two quantized states is located at ~180 meV and ~65 meV for the outer and the inner band respectively, while the Fermi momenta ($k_F$) are at ~0.18 Å$^{-1}$ and ~0.11 Å$^{-1}$. These values are 25% larger than those obtained for the as-grown sample reported in panel f). In addition the bands of the annealed sample are located at ~45meV higher binding energies. All these changes confirm electron doping of the sample due to the increase of V$_{OS}$ by annealing, as observed also elsewhere.[22]

For a direct comparison between the $k_F$ of the annealed and as-grown samples, Figure 1g) reports the momentum distribution curves (MDCs) as extracted at the Fermi level for the two films (red/blue lines in panel e/f respectively), while panel h) shows the Energy Distribution Curves (EDCs) obtained from panel e and f at the two different $k_F$ vectors. In the latter, a broad intense non-dispersive state is found between ~1eV and 2 eV of binding energy (BE), corresponding to localized IG states[13, 23]. In the annealed samples, the spectral weight of IG states is shifted towards the Fermi level; the asymmetric shape of the peak is consistent with the presence of a second in-gap state at lower BE (~1 eV). Such state appears also upon beam irradiation (see supplemental information Figure S3) and its spectroscopic intensity reaches a saturation value. This already suggests that the localised IG states located at ~ 1.3 and 1 eV of BE are related to different types of defects or to different oxygen vacancy sites and that the formation of the shallower defects is favoured by photoirradiation. Similar to the IG states, also the 2DEG intensity increases and saturates, in agreement with previous reports[10, 13]. A further significant spectral change in the two EDCs of panel h) is visible at the Fermi level: a shoulder on the 2DEG peak of the as-grown sample smears out after the annealing, consistently with an increased number of carriers [9, 12]. The find the electron carrier density values for the two samples $n_{2D}^{as\text{-}grown}$~3·10$^{13}$ and $n_{2D}^{annealed}$~5·10$^{13}$ cm$^{-2}$, as calculated via Luttinger's theorem for two dimensional states with spin-degenerate bands[24].

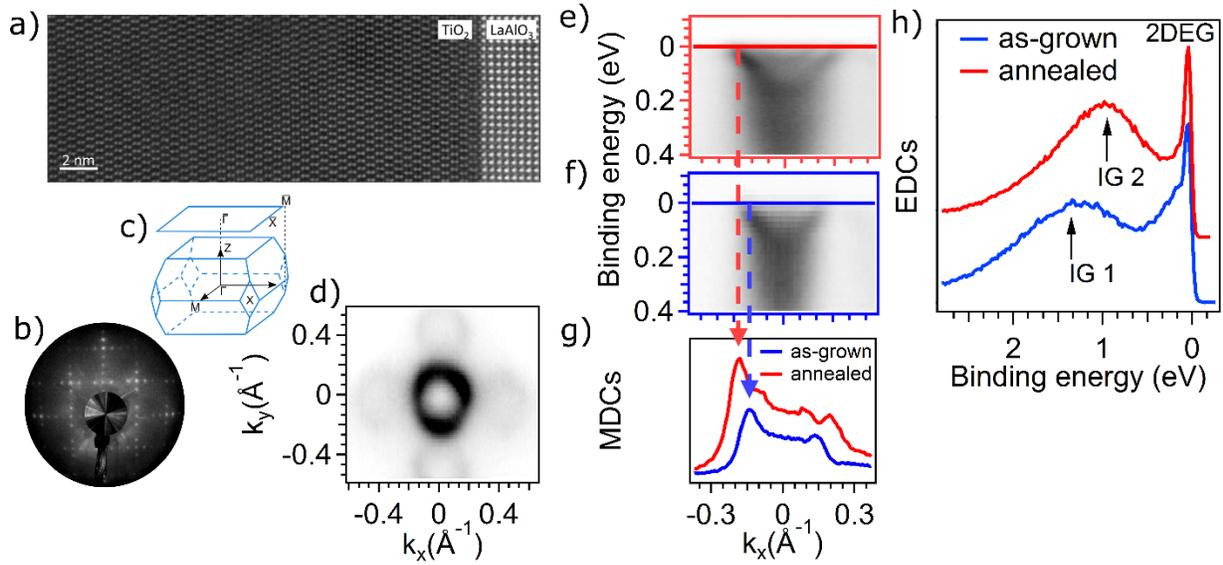

*Figure 1 (a)* Representative high-resolution Z-contrast image of the TiO$_2$/LAO interfacial region. The interface region is atomically sharp and no interdiffusion is observed between film and substrate. The typical dumbbell structure of Ti ions in TiO$_2$ anatase is clearly distinguishable in the film and occurs in the entire film region with no sign of presence of secondary phases; *(b)* LEED pattern (~110eV) showing the (1×4)-(4×1) surface reconstruction of the anatase thin films; *(c)* sketch of the first BZ scheme of anatase; *(d)* Fermi surface contour, measured at 46 eV photon energy, covering the first BZ. It was obtained superimposing the Fermi surfaces measured with different light polarisations (i.e. linear horizontal and vertical, circular right and left). With such a procedure we could compensate the lack of intensity due to symmetry-related selection rules typically occurring for bands of d$_{xy}$ orbital character (see Supplementary Fig. S2). *(e), (f)* ARPES spectra of the metallic state acquired at hv=46 eV photon energy around the $\bar{\Gamma}$ of the second Brillouin zone ($\bar{\Gamma}_{II}$) *(e)* 2DEG of anatase film with high amount of oxygen vacancies after the annealing treatment; *(f)* 2DEG of as-grown film; *(g)* MDCs at the Fermi level , in correspondence of the straight lines in panels (e) and (f)) for the annealed (red) and as-grown (blue) samples respectively, the dashed lines highlight the position of the Fermi momenta k$_F$. *(f)* EDCs extracted at the k$_F$ of the outer band for the annealed (red) and the as-grown (blue) samples.

While some reports suggest that the metallic state has a 3D character[9] a model linking the metallic state to the specific anatase surface arrangement has recently been shown to provide excellent agreement with the experimental data[11]. The 2D nature of the metallic state is also supported by experiments studying both the effect of electron doping through alkaline adsorption and the influence of beam irradiation at the anatase surface[10, 12]. To gain further insight, we have performed soft X-ray ARPES and Resonant PES (RESPES) experiments compensating the production of oxygen vacancies arising from photoirradiation. This has been achieved during the measurements by *in-operando* fluxing molecular oxygen through a metal capillary positioned in the proximity of the sample surface [23, 25, 26]. Figure 2 shows the ARPES spectra for the pristine sample measured with linearly polarised radiation of hν = 120 eV (panel a) and ARPES measured on the very same sample while dosing the surface with oxygen (panel b). In panel c), the MDCs extracted at $E_F$ from panels a) and b) are compared. The value of $k_F$ is reduced from 0.19 Å$^{-1}$ to 0.15 Å$^{-1}$ (~20%) under oxygen dosing, i.e. it shows the opposite trend compared to that observed upon annealing in Figure 1e) and 1f), giving direct evidence of (partial) healing of V$_O$s and consequent reduction of the number of free electron carriers at the surface, in agreement with previous reports[9]. Such a decrease corresponds to a reduction in the carrier density of ~60%, qualitatively consistent with the theoretical picture for subsurface vacancies presented below.

Figure 2d) compares the angle-integrated Density of States (DOS) extracted from panels (a) and (b) with a BE range covering also the IG states. Almost complete suppression of the IG states is observed as soon as oxygen is dosed; the residual IG intensity is little affected by further increase of the oxygen partial pressure (4 times the initial value), dropping below the background signal (it is visible only at resonance, see next section). Conversely, minor changes are observed in the metallic 2DEG intensity upon dosing. Such a distinct behaviour of IG and 2DEG has not been reported before in anatase TiO$_2$, while similar observations exist for LaAlO$_3$/SrTiO$_3$ interfaces[23].

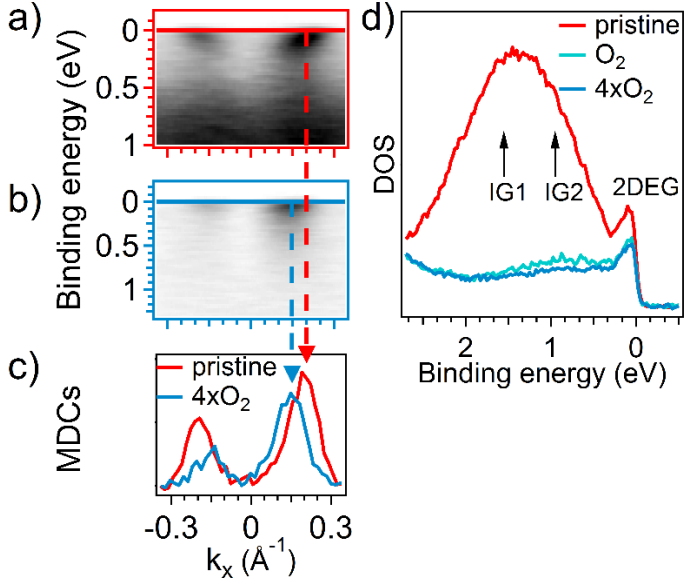

*Figure 2* ARPES spectra of the IG and 2DEG of a TiO$_2$/LaAlO$_3$ sample acquired in the second BZ using hv =120 eV linearly horizontal polarised radiation: **(a)** pristine as-grown samples (base pressure 1x10$^{-10}$ mbar) and **(b)** with molecular dosing (base pressure 4x10$^{-9}$ mbar; partial O$_2$ pressure 4x10$^{-10}$ mbar); **(c)** MDCs extracted at E$_f$ (red and light blue straight lines in panels **(a)** and **(b)** respectively indicate where the MDCs have been extracted) showing a significant difference in k$_F$ (dashed red and blue lines); **(d)** Evolution of the DOS upon oxygen dosing (red line corresponds to partial pressure P = 1 x 10$^{-10}$ mbar, light blue P = 1 x 10$^{-9}$ mbar and dark blue P = 4x10$^{-10}$ mbar), a dramatic reduction of the IG states is observed with only minor changes of the 2DEG.

## III. THEORETICAL STUDIES
### A. COMPUTATIONAL DETAILS

DFT calculations were performed using the Vienna Ab Initio Simulation Package (VASP)[37, 38]. We used the projector augmented-wave (PAW) pseudopotentials to describe the electron-ion interactions and the PBE functional[39] within the generalized gradient approximation (GGA) to treat the exchange-correlation interaction between electrons. The energy cut-off for the expansion of the wave-functions was set to 500 eV. Since GGA always favours delocalized electronic states, selected calculations using the PBE+U method with U = 3.9 eV [40] were also carried out in order to check the robustness of the PBE solutions. Although such a large U value does not accurately describe the thermodynamic properties of TiO$_2$[41] (e.g. it does not reproduce the observed greater stability of subsurface vs surface VOs at the anatase (101) surface22), the results of these calculations largely confirmed the PBE results (see Supplementary Fig. S7). To model the anatase TiO$_2$ (001)-(1 × 4) surface, we used slabs of 8 TiO$_2$ layers with a (3×4) surface supercell for calculations of defect formation energies, in order to minimize interactions between defects in periodic replicas, and slabs of 12 TiO$_2$ layers with a (2×4) surface supercell for calculations of the electronic structure and charge densities. The vacuum region between consecutive slabs was > 12 Å. A 3×2×1 Monkhorst-Pack mesh was used to sample the Brillouin zone. Geometry optimizations were carried out with convergence thresholds of $10^{-4}$ eV and $1\times10^{-2}$ eV/Å for the total energy and the forces on the ions, respectively. Oxygen vacancy formation energies were calculated as E$_{form}$ (VO) = E$_{def}$ − E$_{stoich}$ + 1/2 E$_{tot}$ (O2), where E$_{def}$ and E$_{stoich}$ are the total energies of the reduced (defective) and stoichiometric (defect-free) slabs, respectively, and E$_{tot}$ (O2) is the total energy of the O2 molecule. Reaction pathways were determined using the climbing image nudged elastic band method.[42]

## B. THEORETICAL RESULTS

The observed different dependence of IG and 2DEG states to oxygen dosing suggests that they could be linked to different vacancy sites. To confirm[27] this hypothesis, we performed DFT calculations of surface and subsurface V$_O$s at the reconstructed anatase (001)–(1×4) surface using the well-established model of Ref. [19] (Fig. 3a). From the computed V$_O$ formation energies (**Figure 3**b; see Supplemental information for computational details), it appears that oxygen vacancies are most likely to occur at the surface VO1 ridge site and at the subsurface VO7 site, while other surface sites, notably VO2 and VO3, are energetically unfavourable, thus hindering vacancy migration from VO1 to the subsurface and vice-versa. Importantly, Figs. 3c and 3d, and Supplemental Figs. S4 and S5 further show that the character of the electronic states is very different for surface and subsurface V$_O$s. The two excess electrons are well localized on the Ti atoms adjacent to the vacant site and form deep energy levels in the band gap in the case of VO1 (in agreement with previous calculations by Shi et al. [28]) and other surface sites. In contrast, the electrons are delocalized over few (001) planes around the vacancy and form shallow energy levels at the bottom of the conduction band in the case of subsurface VO6, VO7 and deeper V$_O$s (consistent with Ref. [29]).

The surface potential in the absence/presence of V$_{OS}$ was computed from the energies of the semi-core Ti 3s levels in different layers of the slab (Supplementary Figs. S6 and S7). As shown in Figure 3e), the potential becomes repulsive near the surface of a defect-free slab[11, 29], and this effect is further enhanced in the presence of VO1 (green line). In contrast, VO7 (red line) induces an attractive potential well (depth ~ 0.2 eV) that confines the excess electron states in the neighbouring Ti layers. Note that, unlike in previous modelling studies[11], this confining potential emerges naturally in our calculations for the reduced slabs.

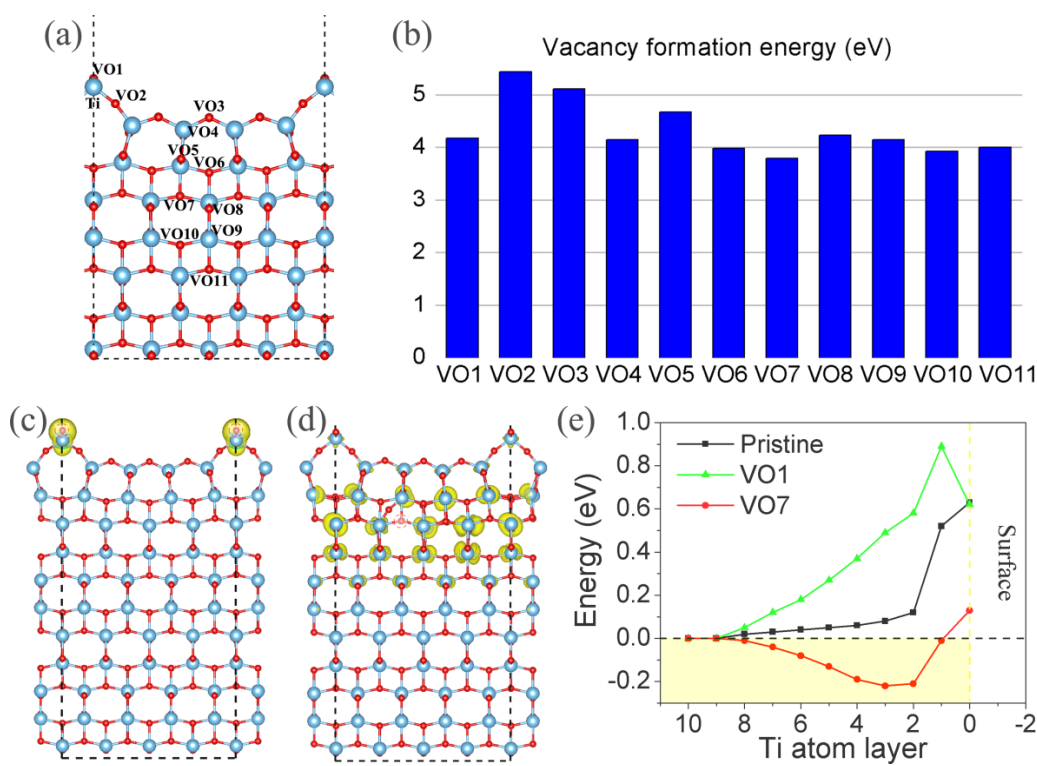

*Figure 3 (a) Side view of the reconstructed anatase TiO$_2$(001)-1×4 slab model (8 TiO$_2$ layers, (3×4) surface supercell) used to calculate O-vacancy formation energies at different surface and subsurface sites, as indicated; Ti and O atoms are light blue and red, respectively. Ridges exposing twofold coordinated oxygen (O$_{2c}$) and fourfold Ti atoms (Ti$_{4c}$) are separated by valleys exposing O$_{2c}$ and fivefold Ti (Ti$_{5c}$) atoms with a (1×4) periodicity. (b) V$_O$ formation energies (eV; blue bars) at different surface and subsurface oxygen sites computed using DFT-PBE; (c,d) Charge density contours of the excess electron states induced by VO1 and VO7, respectively; the vacancy positions are indicated by dashed red circles; dashed black lines show the unit cell used for the calculations;*

*(e) Surface potential from the shift of the Ti 3s peak in the different layers of the pristine and reduced slabs. The yellow shading highlights the region of negative (attractive) surface potential.*

To model the effect of oxygen dosing, we considered the adsorption of an $O_2$ molecule on the anatase surface with a surface or subsurface $V_O$ (**Figure 4**). $O_2$ adsorption on $TiO_2$ is known to involve the transfer of excess electrons from the oxide to the molecule [1, 3, 27, 30]. In the presence of a VO1, $O_2$ undergoes a strongly exothermic and barrier-less adsorption at the vacancy site (see Supplemental Figure S8), which results in the formation of a bridging peroxide ($O_2^{2-}$) at the ridge, denoted $(O_2)_O$ in Figure 4. Since the two excess electrons of VO1 are both transferred to the adsorbed species, no excess electron remains in $TiO_2$, consistent with the strong reduction of the IG signal observed in ARPES when exposing the surface to $O_2$.

A different picture holds for the adsorption of $O_2$ on a surface with subsurface $V_{OS}$. In this case, only one of the two excess electrons of the vacancy transfers to the $O_2$ molecule [1, 3, 30], thus resulting in the formation of an adsorbed superoxide ($O_2^-$), denoted $O_2^*$ in Figure 4. This negatively charged species has an attractive interaction with the positively charged subsurface vacancy, so that migration of $V_O$ toward the surface would be energetically favourable (Figure 4a). However, at variance with what was found for anatase (101)[27], the energy barrier for subsurface → surface migration of the $V_O$ is quite high at the anatase (001) surface[31] (see Figure 4b) for the VO4→ VO3 migration step), at least for the $O_2$ concentration considered here. Thus, the $V_O$ remains subsurface, the adsorbed $O_2$ remains a superoxide, and one of the two excess electrons remains in $TiO_2$. This explains the persistence of the 2DEG signal as well as the decrease in the number of carriers observed under oxygen dosing.

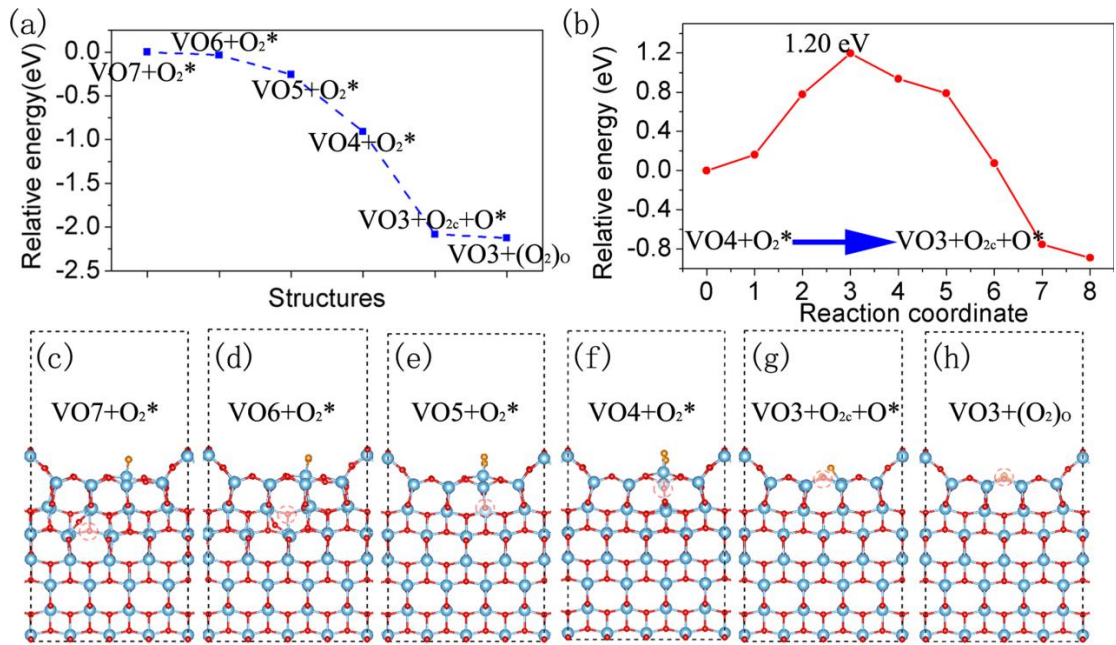

*Figure 4 (a) $O_2$ adsorption energy as a function of the subsurface (VO4 – VO7) or surface (VO3) oxygen vacancy location. Relevant structures with a subsurface $V_O$, denoted $VOn+O_2*$ (n=4 -7), are shown in panels (c-f). For VO3, two nearly degenerate structures are present, as shown in panels (g-h), where O\* indicates an oxygen adatom and $(O_2)_O$ a bridging peroxide replacing an $O_{2c}$. The energy zero corresponds to the adsorption energy of $VO7+O_2*$. (b) Energy barrier for the diffusion of an O-vacancy from VO4 to VO3 in the presence of adsorbed oxygen. (c-h) Atomic geometries of adsorbed $O_2$ on reduced anatase (001) with a subsurface (VO4-VO7) or surface (VO3) oxygen vacancy, as described in (a). Ti atoms are blue, O atoms are red, adsorbed $O_2$ is orange; dashed red circles indicate the positions of the vacant sites.*

## IV.  EXPERIMENTAL RESULTS 2 – RESONANT PHOTOEMISSION (RESPES)

To verify the scenario provided by the DFT results, we performed resonant photoemission (RESPES) measurements at the Ti $L_{2,3}$ edges. As both IG and 2DEG arise from Ti 3d states[11,12], RESPES provides additional information by exploiting the energy shift between the core levels of titanium atoms with different oxidation states[31]. Figure 5a) shows the X-ray absorption spectra (XAS) across the Ti $L_3$ edge from an as-grown sample (red curve) and under oxygen dosing (dark blue curve). In both XAS spectra the more prominent peaks are located at typical energies for titanium $Ti^{4+}$ in crystalline anatase [32, 33,34]. Upon oxygen dosing, the spectral intensity is lowered in the pre-edge as well as in the valleys at ~459 eV and ~462.5 eV, which correspond to spectral lines of $Ti^{3+}$ [33]. As found in similar systems, e.g. rutile $TiO_2$[35] and $SrTiO_3$[36], the observed changes can be directly linked to the number of oxygen vacancies.

The angle integrated photoemission EDCs, displayed as a color map in Figure 5b), indicate that the IG and 2DEG states resonate at different photon energies (more detailed results are displayed in Supplemental Figure S9). The IG is peaked at an energy corresponding to a $Ti^{3+}$ oxidations state, i.e. the valley at ~459.3 eV (roughly 0.8 eV away from the resonance of the 2DEG states), in agreement with previous results[20, 23]. Conversely, the 2DEG follows the same trend of the XAS, with maximum intensity located around the $L_3$-$e_g$ doublet at ~460.5 eV (characteristic of the stoichiometric $Ti^{4+}$). Similarly, above 462 eV (i.e. in the valley before the $L_2$-edge) the IG intensity rises first. Altogether, these findings provide evidence that: i) the IGs are strongly localised on Ti atoms close to vacancy sites ($Ti^{3+}$) and ii) the 2DEG wavefunction is delocalised over many Ti sites, consistent with the theoretical results discussed above (see Figure 3c) and d)).

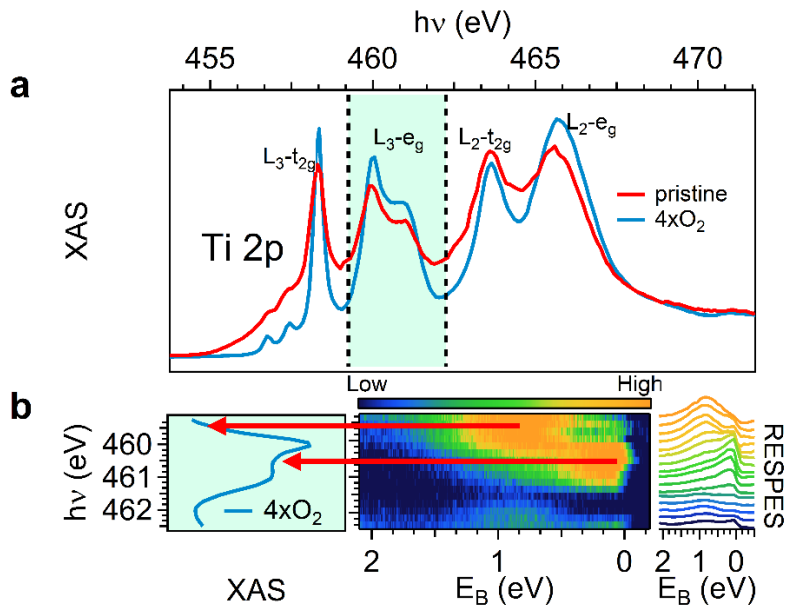

*Figure 5 (a) Effect of $O_2$ dosing on the absorption spectrum (total yield) at Ti $L_{2,3}$ edges of anatase $TiO_2$ film grown on Nb-doped $SrTiO_3$, measured at 90 K. Pristine sample (red) refers to base pressure in the chamber of $1x10^{-10}$ mbar, while the dark blue curve corresponds to the highest oxygen partial pressure of the present work ($4x10^{-9}$ mbar). The four main peaks (located at approximate excitation energies of 458.3 eV, 460.5 eV, 463.5 eV, and 465.5 eV excitation energies) are ascribable to the combination of spin-orbit splitting of the initial states ($L_3$-$L_2$) and crystal field* *splitting of the d orbitals in the final state ($t_{2g}$-$e_g$) for Ti atoms in 4+ oxidation state. The additional splitting of the $L_3$-$e_g$ peak (~ 460.5 eV) is the fingerprint of the anatase phase arising from distortion of the ideal octahedron around Ti atom and long-range effects.[32, 33] **(b)** ResPES (circular polarization, T= 90 K) acquired in the second BZ of a $TiO_2$/$LaAlO_3$ sample upon oxygen dosing (P = 1x10$^{-9}$ mbar) to better decouple the IGs and the 2DEG signals (Supplemental Fig. S9). The colour map (central panel) displays the momentum-integrated photoemission intensity as a function of the photon energy and of the electron binding energy. The resonant DOS curves (right panel) show the different resonating energies for the IG and the 2DEG states (red arrows).*

## V. SUMMARY AND PERSPECTIVE

In summary, our results reveal distinct behaviours of localized and delocalized states induced by oxygen vacancies at the surface of anatase $TiO_2$. Due to their different spatial locations and the kinetics of defect diffusion in anatase, the 2D delocalized states are much more robust than the localised in-gap states when exposed to molecular oxygen in a wide range of pressures. This robustness of the delocalized states is an important feature that could be exploited for different applications, e.g. to tune the electronic structure of $TiO_2$ in engineered interfaces and heterostructures, or to precisely control the concentration of charge carriers in photo-sensitive devices.

Our findings unveil a new and relevant aspect of the surface chemistry of $TiO_2$, providing a pathway to tailored performances of future devices.

ACKNOWLEDGMENT

This work has been partly performed in the framework of the nanoscience foundry and fine analysis (NFFA-MIUR Italy) facility. We thank Fabio Miletto Granozio (Spin-CNR) and Ralph Claessen (Univ. Wurzburg) for useful discussions. A.S. acknowledges the support of DoE-BES, Division of Chemical Sciences, Geo sciences and Biosciences under Award DE-SC0007347. Z. T. was supported by the National Natural Science Foundation of China (No. 51602092), the Science and Technology Innovation Project Foundation of Hunan Province (No. 2018RS3103). GD acknowledges the financial support from Slovenian Research Agency (P2-0393). E. Cociancich from CNR-IOM and A. Di Cristoforo from Università Politecnica delle Marche are gratefully acknowledged for the support in the TEM specimen preparation.

REFERENCES

1. T. L. Thompson, J. T. Yates, *Surface science studies of the photoactivation of $TiO_2$-new photochemical processes*, Chemical Reviews **106** (10), 4428-4453 (2006).

2. A. Fujishima, X. Zhang, D. A. Tryk, *$TiO_2$ photocatalysis and related surface phenomena*. Surface Science Reports **63** (12), 515-582 (2008).

3. U. Diebold, *The surface science of titanium dioxide*, Surface Science Reports **48** (5), 53-229 (2003).

4. C. Di Valentin, G. Pacchioni, A. Selloni, *Reduced and n-Type Doped TiO2: Nature of Ti3+ Species*, Journal of Physical Chemistry C **113** (48), 20543-20552 (2009).

5. M. A. Henderson, *A surface science perspective on $TiO_2$ photocatalysis*, Surface Science Reports 66 (6-7), 185-297 (2011).

6. S. X. Zhang, D. C. Kundaliya, W. Yu, S. Dhar, S. Y. Young, L.G. Salamanca-Riba, S. B. Ogale, R. D. Vispute, T. Venkatesan, *Niobium doped TiO2: Intrinsic transparent metallic anatase versus highly resistive rutile phase,* Journal of Applied Physics **102** (1), 013701 (2007).

7. J. Bai, B. Zhou, *Titanium Dioxide Nanomaterials for Sensor Applications*, Chemical Reviews **114** (19), 10131-10176 (2014).

8. A. G. Thomas, W. R. Flavell, A. K. Mallick, A. R. Kumarasinghe, D. Tsoutsou, N. Khan, C. Chatwin, S. Rayner, G. C. Smith, R. L. Stockbauer, S. Warren, T. K. Johal, S. Patel, D. Holland, A. Taleb, F. Wiame, *Comparison of the electronic structure of anatase and rutile TiO2 single-crystal surfaces using resonant photoemission and x-ray absorption* *spectroscopy,* Physical Review B (Condensed Matter and Materials Physics) **75** (3), 035105 (2007).

9. S. Moser, L. Moreschini, J. Jacimovic, O. S. Barisic, H. Berger, A. Magrez, Y. J. Chang, K. S. Kim, A. Bostwick, E. Rotenberg, L. Forró, M. Grioni, *Tunable Polaronic Conduction in Anatase TiO2*, Physical Review Letters **110** (19), 196403 (2013).

10. T. C. Rödel, F. Fortuna, F. Bertran, M. Gabay, M. J. Rozenberg, A. F. Santander-Syro, P. Le Fèvre, *Engineering two-dimensional electron gases at the (001) and (101) surfaces of TiO2 anatase using light,* Physical Review B **92**, 041106(R) (2015).

11. Z. Wang, Z. Zhong, S. McKeown Walker, Z. Ristic, J. Z. Ma, F. Y. Bruno, S. Riccò, G. Sangiovanni, G. Eres, N. C. Plumb, L. Patthey, M. Shi, J. Mesot, F. Baumberger, M. Radovic, *Atomically Precise Lateral Modulation of a Two-Dimensional Electron Liquid in Anatase TiO2 Thin Films*, Nano Letters **17** (4), 2561-2567 (2017).

12. R. Yukawa, M. Minohara, D. Shiga, M. Kitamura, T. Mitsuhashi, M. Kobayashi, K. Horiba, H. Kumigashira, *Control of two-dimensional electronic states at anatase TiO2(001) surface by K adsorption,* Physical Review B **97** (16), 165428 (2018).

13. Y. Aiura, K. Ozawa, E. F. Schwier, K. Shimada, K. Mase, *Competition between Itineracy and Localization of Electrons Doped into the Near-Surface Region of Anatase TiO2,* The Journal of Physical Chemistry C **122** (34), 19661-19669 (2018).

14. A. Ohtomo, H. Y. Hwang, *A high-mobility electron gas at the LaAlO3/SrTiO3 heterointerface,* Nature **427**, 423 (2004).

15. A. F. Santander-Syro, O. Copie, T. Kondo, F. Fortuna, S. Pailhès, R. Weht, X. G. Qiu, F. Bertran, A. Nicolaou, A. Taleb-Ibrahimi, P. Le Fèvre, G. Herranz, M. Bibes, N. Reyren, Y. Apertet, P. Lecoeur, A. Barthélémy, M. J. Rozenberg, *Two-dimensional electron gas with universal subbands at the surface of SrTiO3,* Nature **469**, 189 (2011).

16. P. D. C. King, R. H. He, T. Eknapakul, P. Buaphet, S. K. Mo, Y. Kaneko, S. Harashima, Y. Hikita, M. S. Bahramy, C. Bell, Z.Hussain, Y. Tokura, Z. X.Shen, H. Y. Hwang, F. Baumberger, W. Meevasana, *Subband Structure of a Two-Dimensional Electron Gas Formed at the Polar Surface of the Strong Spin-Orbit Perovskite KTaO3*, Physical Review Letters **108** (11), 117602 (2012).

17. R. Hengerer, B. Bolliger, M. Erbudak, M. Grätzel, *Structure and stability of the anatase TiO2 (101) and (001) surfaces*, Surface Science **460** (1), 162-169 (2000).

18. G. S. Herman, M. R. Sievers, Y. Gao, *Structure Determination of the Two-Domain ( 1 x 4) Anatase TiO2(001) Surface*, Physical Review Letters **84**, 3354 (2000).

19. M. Lazzeri, A. Selloni, *Stress-Driven Reconstruction of an Oxide Surface: The Anatase TiO2(001) (1x4) Surface*, Physical Review Letters **87** (26), 266105 (2001).

20. B. Gobaut, P. Orgiani, A. Sambri, E. Di Gennaro, C. Aruta, F. Borgatti, V. Lollobrigida, D. Céolin, J.-P. Rueff, R. Ciancio, C. Bigi, P.K. Das, J. Fujii, D. Krizmancic, P. Torelli, I. Vobornik, G. Rossi, F. Miletto Granozio, U. Scotti di Uccio, G. Panaccione, *Role of Oxygen*

*Deposition Pressure in the Formation of Ti Defect States in TiO2(001) Anatase Thin Films*, ACS Applied Materials & Interfaces **9** (27), 23099-23106 (2017).

21. P. D. C. King, S. McKeown Walker, A. Tamai, A. de la Torre, T. Eknapakul, P. Buaphet, S. K. Mo, W. Meevasana, M. S. Bahramy, F. Baumberger, *Quasiparticle dynamics and spin–orbital texture of the SrTiO3 two-dimensional electron gas*, Nature Communications **5**, 3414 (2014).

22. P. Scheiber, M. Fidler, O. Dulub, M. Schmid, U. Diebold, W. Hou, U. Aschauer, A. Selloni, *(Sub)Surface Mobility of Oxygen Vacancies at the TiO2 Anatase (101) Surface*, Physical Review Letters **109** (13), 136103 (2012).

23. J. Gabel, M. Zapf, P. Scheiderer, P. Schütz, L. Dudy, M. Stübinger, C. Schlueter, T. L. Lee, M. Sing, R. Claessen, *Disentangling specific versus generic doping mechanisms in oxide heterointerfaces*, Physical Review B **95** (19), 195109 (2017).

24. J. M. Luttinger, *Fermi Surface and Some Simple Equilibrium Properties of a System of Interacting Fermions*, Physical Review **119** (4), 1153-1163 (1960).

25. L. Dudy, M. Sing, P. Scheiderer, J. D. Denlinger, P. Schütz, J. Gabel, M. Buchwald, C. Schlueter, T.-L. Lee, R. Claessen, *In Situ Control of Separate Electronic Phases on SrTiO3 Surfaces by Oxygen Dosing*, Advanced Materials **28** (34), 7443-7449 (2016).

26. P. Scheiderer, M. Schmitt, J. Gabel, M. Zapf, M. Stübinger, P. Schütz, L. Dudy, C. Schlueter, T.-L. Lee, M. Sing, R. Claessen, *Tailoring Materials for Mottronics: Excess Oxygen Doping of a Prototypical Mott Insulator*, Advanced Materials **30** (25), 1706708 (2018).

27. M. Setvín, U. Aschauer, P. Scheiber, Y.-F. Li, W. Hou, M. Schmid, A. Selloni, U. Diebold, *Reaction of O2 with Subsurface Oxygen Vacancies on TiO2 Anatase (101)*, Science **341** (6149), 988 (2013).

28. Y. Shi, H. Sun, M. C. Nguyen, C. Wang, K. Ho, W. A. Saidi, J. Zhao, *Structures of defects on anatase TiO2(001) surfaces*, Nanoscale **9** (32), 11553-11565 (2017).

29. S. Selcuk, A. Selloni, *Facet-dependent trapping and dynamics of excess electrons at anatase TiO2 surfaces and aqueous interfaces*, Nature Materials **15**, 1107 (2016).

30. Y.-F. Li, A. Selloni, *Theoretical Study of Interfacial Electron Transfer from Reduced Anatase TiO2(101) to Adsorbed O2*, Journal of the American Chemical Society **135** (24), 9195-9199 (2013).

31. S. Selcuk, X. Zhao, A. Selloni, *Structural evolution of titanium dioxide during reduction in high-pressure hydrogen*, Nature Materials **17** (10), 923-928 (2018).

32. P. Krüger, *Multichannel multiple scattering calculation of L2,3-edge spectra of TiO2 and SrTiO3: Importance of multiplet coupling and band structure*, Physical Review B 81 (12), 125121 (2010).

33. F. M. F. de Groot; M. O. Figueiredo, M. J. Basto, M. Abbate, H. Petersen, J. C. Fuggle, *2p X-ray absorption of titanium in minerals*, Physics and Chemistry of Minerals **19** (3), 140-147 (1992).

34. S. O. Kucheyev, T. van Buuren, T. F. Baumann, J. H. Satcher, T. M. Willey, R. W. Meulenberg, T. E. Felter, J. F. Poco, S. A. Gammon, L. J. Terminello, *Electronic structure of titania aerogels from soft x-ray absorption spectroscopy*, Physical Review B **69** (24), 245102 (2004).

35. P. Le Fèvre, J. Danger, H. Magnan, D. Chandesris, J. Jupille, S. Bourgeois, M. A. Arrio, R. Gotter, A. Verdini, A. Morgante, *Stoichiometry-related Auger lineshapes in titanium oxides: Influence of valence-band profile and of Coster-Kronig processes*, Physical Review B **69** (15), 155421 (2004).

36. C. Chen, J. Avila, E. Frantzeskakis, A. Levy, M. C. Asensio, *Observation of a two-dimensional liquid of Fröhlich polarons at the bare SrTiO3 surface*, Nature Communications **6**, 8585 (2015).

37. G. Kresse, J. Furthmüller, *Efficient iterative schemes for ab initio total-energy calculations using a plane-wave basis set*, Phys. Rev. B **54** (16), 11169-11186 (1996).

38. G. Kresse, J. Furthmüller, *Efficiency of ab-initio total energy calculations for metals and semiconductors using a plane-wave basis set*, Comput. Mater. Sci. **6** (1), 15 (1996).

39. J. P. Perdew, K. Burke, M. Ernzerhof, *Generalized Gradient Approximation Made Simple*, Physical Review Letters **77** (18), 3865-3868 (1996).

40. M. Setvin, C. Franchini, X. Hao, M. Schmid, A. Janotti, M. Kaltak, C. G. Van de Walle, G. Kresse, U. Diebold, *Direct View at Excess Electrons in TiO2 Rutile and Anatase*, Physical Review Letters **113** (8), 086402 (2014).

41. Z. Hu, H. Metiu, *Choice of U for DFT+U Calculations for Titanium Oxides*, The Journal of Physical Chemistry C **115** (13), 5841-5845 (2011).

42. G. Henkelman, B. P. Uberuaga, H. Jónsson, *A climbing image nudged elastic band method for finding saddle points and minimum energy paths*, The Journal of Chemical Physics **113** (22), 9901-9904 (2000).

# Supplemental Information

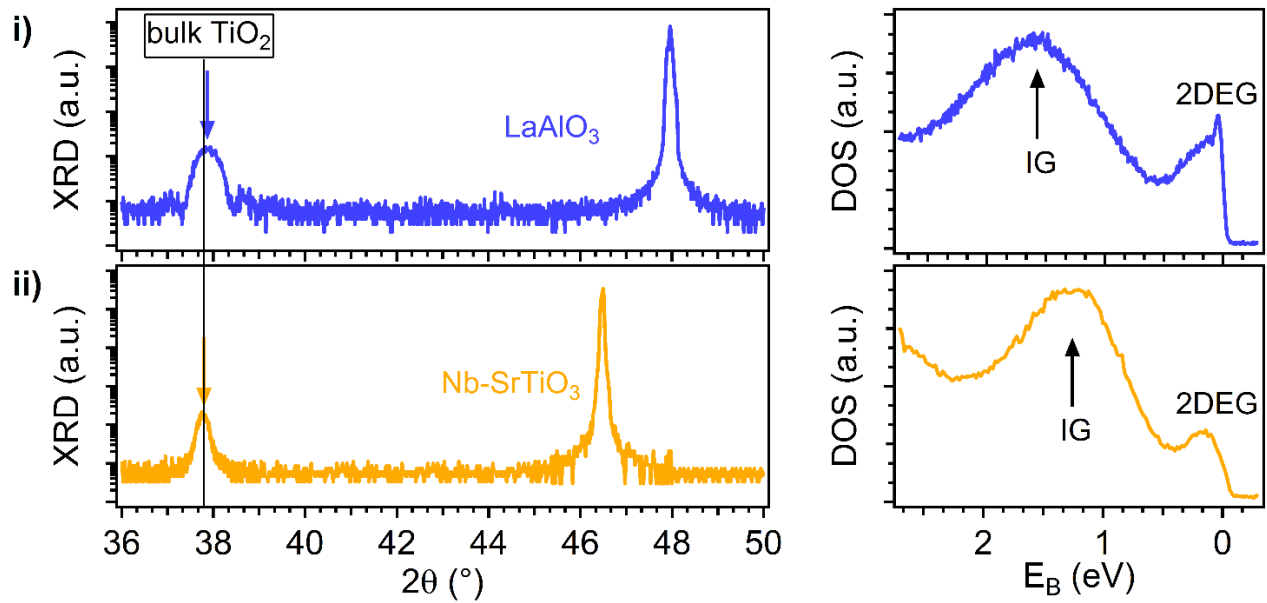

**Supplemental Figure S1**. Structural and electronic characterization of anatase TiO₂ thin films grown on different substrates: i) strain-less TiO₂ films grown on (001) LaAlO₃ (LAO), and ii) fully relaxed films on Nb-doped (1%wt) (001) SrTiO₃ (STO), characterized by a large in-plane lattice mismatch between film and substrate (-3%). <u>Left</u>: θ-2θ high-resolution scans of (004) Bragg reflection of the TiO₂ films, along with the expected bulk TiO₂ peak position. The high structural order of the films is verified by well-defined interference fringes around the Bragg peak, indicating homogeneous orientation, and by the value of the out-of-plane lattice parameter. The large lattice mismatch between TiO2 and SrTiO3 in-plane lattice parameters results into a relaxed growth of the films with lower structural quality, as demonstrated by the absence of interference fringes around the diffraction peak. The calculated out-of-plane lattice parameters for the different TiO₂ films are 9.49 Å (on LAO) and 9.50 Å (on STO), i.e. within 2-sigma of the 9.51 Å bulk value. The overall structural quality of the resulting TiO₂ samples does not affect the electronic properties of the 2DEG. <u>Right</u>: Density of States (DOS) measured at 46 eV photon energy. In both films one observes a stabilization of the 2DEG, while in-gap states are located at different binding energies (~1.8eV for LAO, and ~1.3eV for STO), probably because of differences introduced by the substrate (e.g. strain mechanism/dislocations) into the TiO2 films. Detailed analysis of the energy position of the in-gap states vs. substrate is beyond the scope of the present manuscript.

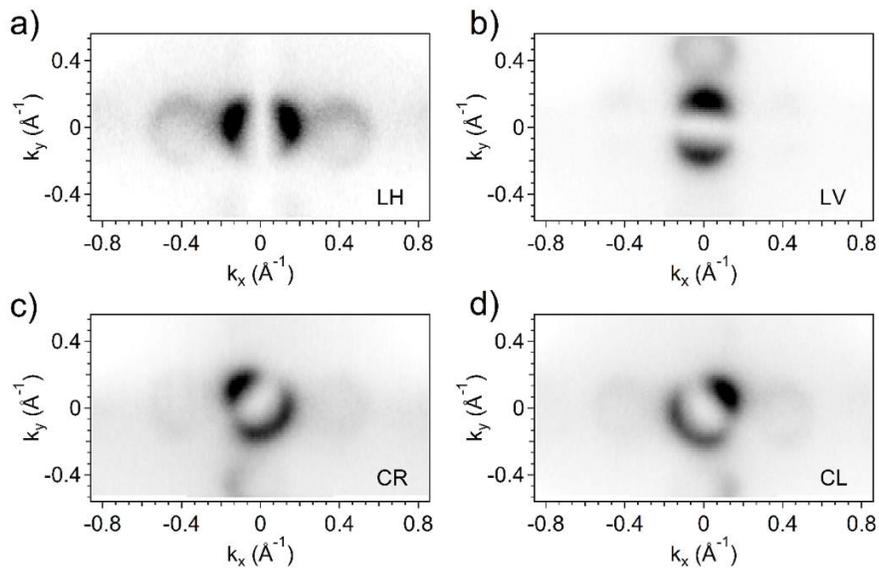

**Supplemental Figure S2** Fermi surface of anatase in the first Brillouin zone (raw data) acquired at 46 eV photon energy as a function of light polarisation: (a) linear horizontal, (b) linear vertical, (c) circular right (**d**) circular left. Fig. 1.d of the main text shows the sum of all the four polarizations; such a procedure has been adopted to compensate the lack of intensity due to symmetry-related issues. By comparing the different panels of the present figure, one can indeed notice gaps in the photoemitted intensity, which rely on the polarisation of the beam impinging on the sample. In particular, these symmetry-related selection rules indicate that the 2DEG state of anatase has a $d_{xy}$ orbital character.[1-4]

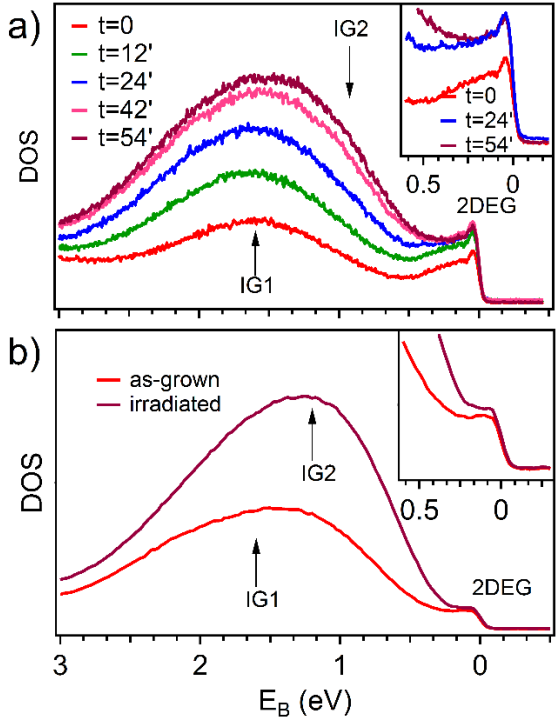

**Supplemental Figure S3.** DOS of anatase films deposited on LAO and measured at (**a**) APE beamline (Elettra; hv=46 eV) and (**b**) I09 beamline (Diamond Light Source; hv=120 eV); in both panels, arrows indicate the position of the IGs for the as-grown and annealed samples, as discussed in the main text. The effects on the electronic properties of intense beam irradiation at the surface of oxides including $TiO_2$ have been widely investigated[5-9]. **Panel a** reports the evolution of the spectral intensity vs. time. The DOS of the pristine surface (red curve) displays in-gap (IG) states at ~1.6eV BE. Under beam irradiation, the intensity of the DOS in the in-gap region increases and stabilises after roughly 1 hour of beam exposure. The metallic 2DEG intensity also increases and saturates after about 25'-30' of beam irradiation. These findings agree with the effects of annealing discussed in the main text and with previous results[5,6], and are consistent with the evolution vs. time observed at Diamond Light Source. The asymmetric shape of the peak in both curves of **panel b** indicates the presence of at least a second, distinct, in-gap state. Similar to changes observed upon annealing, photo-irradiation favours the formation of an IG located at shallower binding energies (i.e. ~1eV BE). This may indicate that the localised states are related to two inequivalent oxygen vacancy sites and that the formation of the latter is more favourable under the beam.

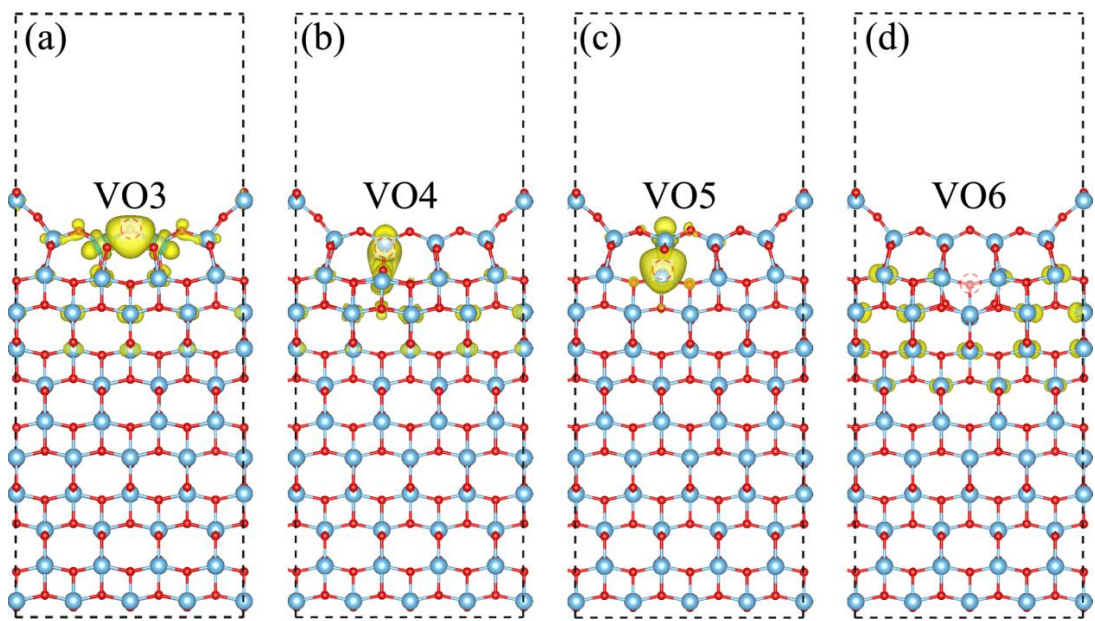

**Supplemental Figure S4** Charge density contours (in yellow) of the excess electron states associated with a few surface and subsurface oxygen vacancies. V$_{OS}$ are labelled as in Fig. 3 of the main text. Red dashed circles indicate the vacancy site.

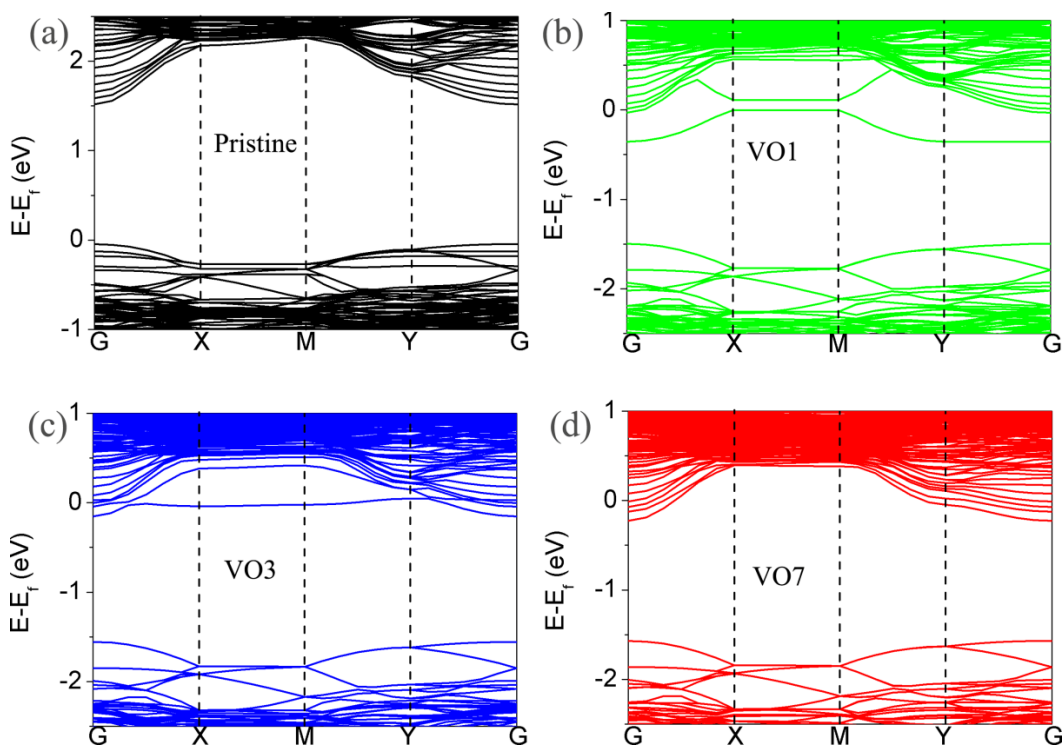

**Supplemental Figure S5** Computed PBE band structure along representative symmetry directions for: **(a)** pristine anatase (001) - 1× 4; **(b,c,d)** reduced surface with an oxygen vacancy at a surface

(VO1, VO3) or subsurface (VO7) site. Following the charge distribution shown in Fig.3 of the main text, VO7 displays clear dispersive features, while VO3 and VO1 tends to localize, with higher effective mass (Supplementary Fig. S4 (for VO3)). On the vertical axes, energies are referred to the Fermi energy of each system. The underestimate of the band gap is a well-known feature of DFT.

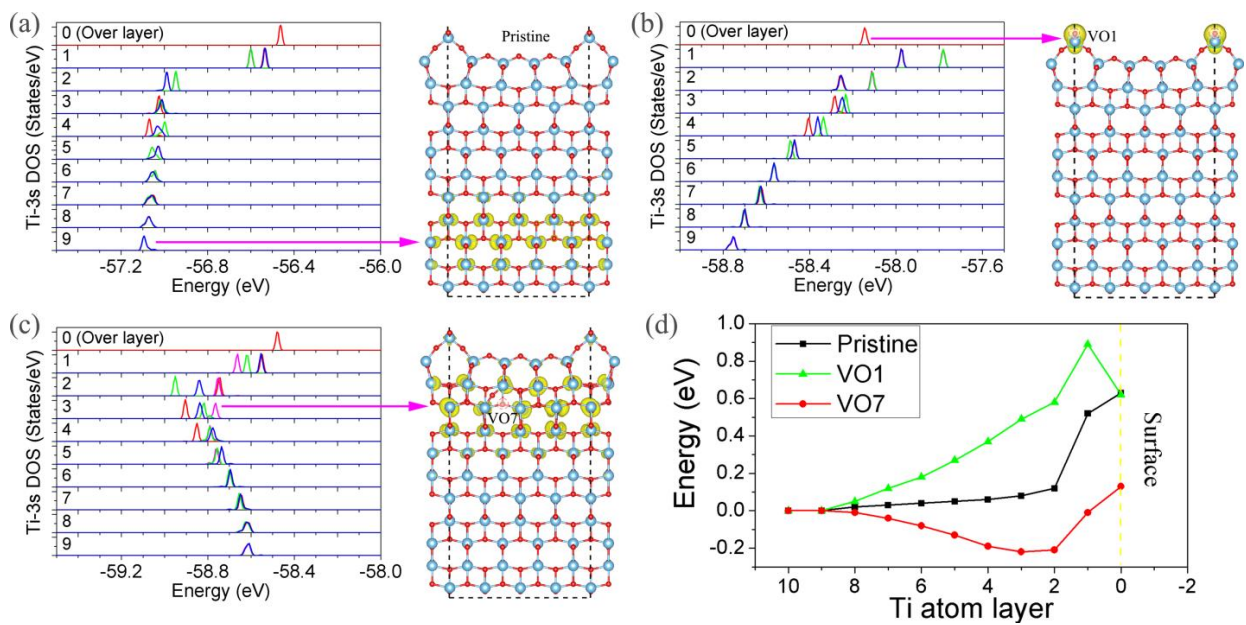

**Supplemental Figure S6** – Computed PBE layer-resolved Ti-3s Density of States and charge density contours for **(a)** stoichiometric anatase (001) – (1× 4) and Ti-3d states at the bottom of the conduction band; **(b)** reduced surface with an oxygen vacancy at VO1 and corresponding excess electron states; **(c)** same as in (b), but for VO7. Layer 0 corresponds to the ridge on the reconstructed surface, layer 1 to the surface, and so forth, as indicated. **(d)** Surface potential from the shift of the Ti 3s peak in the different layers of the pristine and reduced slabs (same plot shown in Fig. 3e).

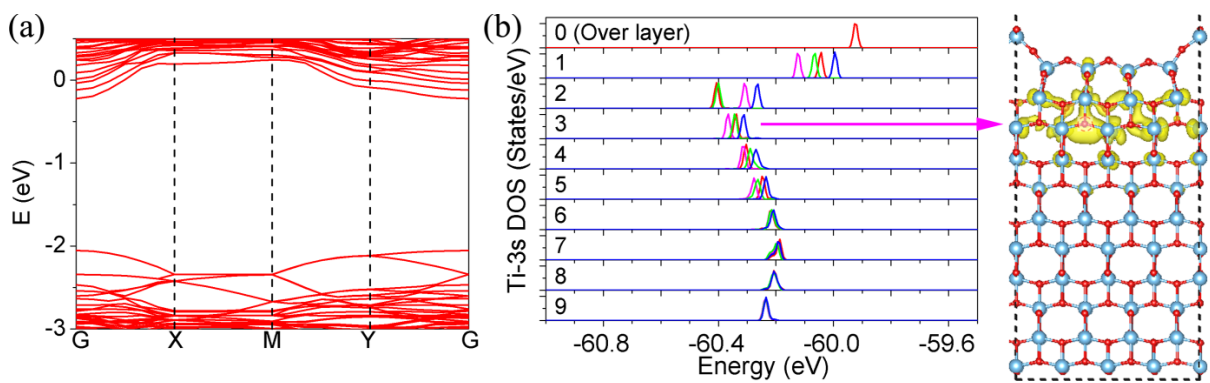

**Supplemental Figure S7** – DFT+U (U=3.9 eV) calculations for reduced anatase (001) – (1× 4) with a subsurface oxygen vacancy at VO7: (a) Band structure; (b) layer-resolved Ti-3s Density of States and defect state (yellow contours). The energy zero corresponds to the Fermi energy. Comparison with the corresponding PBE results in Supplementary Figs. S5d and S6c shows that the main characteristics of the 2DEG (energy bands, confining surface potential, delocalization) are maintained even when a large U value is used.

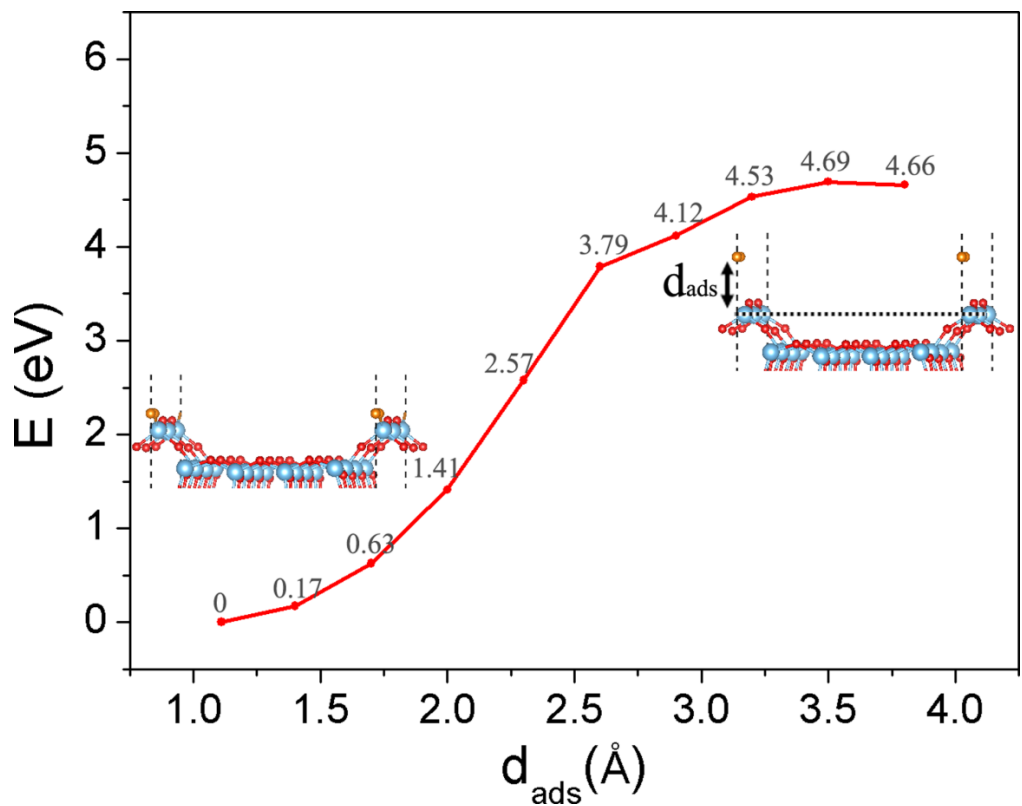

**Supplemental Figure S8.** Potential energy profile as a function of the average Ti-O distance ($d_{ads}$) between the oxygens of an $O_2$ molecule approaching the surface and the Ti atoms on the ridge of the anatase (001) - 1×4 surface with an oxygen vacancy at VO1 (see inset on the right). The energy zero corresponds to the molecule adsorbed at the vacancy site in the form of a bridging peroxo $(O_2)_O$ species (see inset on the left). Ti atoms are light blue, O atoms of $TiO_2$ are red, O atoms of $O_2$ are orange.

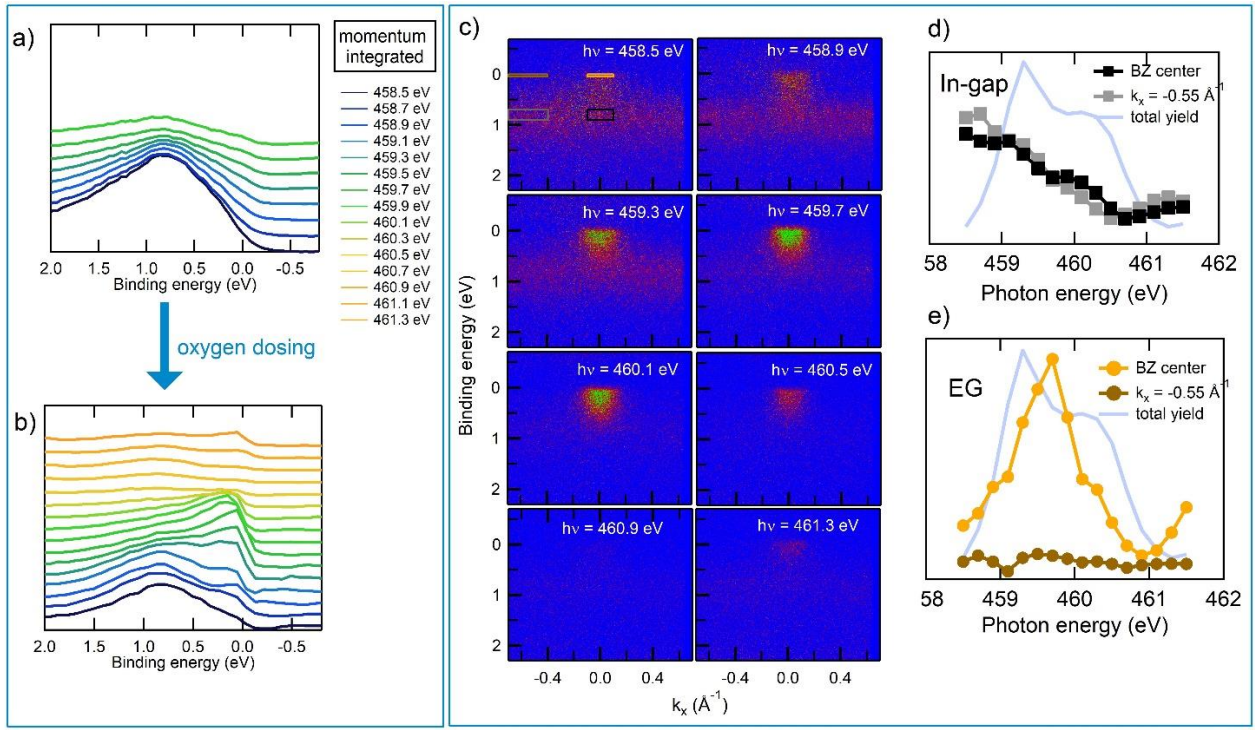

**Supplemental Figure S9**. (**a**), (**b**) DOS evolution at the Fermi level as a function of the photon energy before and during oxygen dosing (spectra in **panel b** are the same as those displayed in the bottom centre and bottom right panels of Fig. 3b in the main text). The IG states are strongly suppressed with oxygen dosing, yet both their BE and their resonating behavior are not affected. The signal of the 2DEG is much clearer in panel b), due to the strong suppression of the IG intensity upon oxygen dosing. (**c**) Resonant angle-resolved-photoemission (ResARPES) spectra acquired in the 2BZ across the Ti $L_3$-$e_g$ absorption edge, measured while dosing oxygen. The in-gap state is well visible as non-dispersive red broad line which resonates at the lowest photon energy (i.e. at 458.5eV), whereas the electron pocket at Fermi is localized in the BZ centre, more intense in the middle of the photon energy scan (459.7eV). (**d**), (**e**) Intensity of the IG and 2DEG states across the Ti $L_3$-$e_g$ absorption edge respectively. Each square (dot) in **panel d** (**panel e**) corresponds to ARPES spectra acquired at different photon energy. The coloured rectangles in the top-left panel of **panel c** mark the binding energy/momentum region where the spectral intensity has been integrated. For both states two regions of momentum space were selected: the projected 2BZ centre ($\bar{\Gamma}_1$) and a low symmetry momentum region. For the IG state, these correspond to the black and grey rectangles, respectively. Similarly, the spectral intensity at the Fermi level has been integrated in the same momentum ranges (i.e. centre of 2BZ (orange) and outside (brown)). As expected, the in-gap state follows the same trend in both k regions, characteristic of a localized state, while the 2DEG state disperses having zero intensity outside the centre (i.e. beyond 0.2 Å$^{-1}$ from $\bar{\Gamma}_1$). In both panels the total electron yield acquired during the measurement is showed for comparison. As in the case of the angle integrated spectra, it is readily observable that the 2DEG state resonates with the L$_3$-e$_g$ doublet, whereas the in-gap state has Ti$^{3+}$ character, resonating in the valley at lower photon energy.

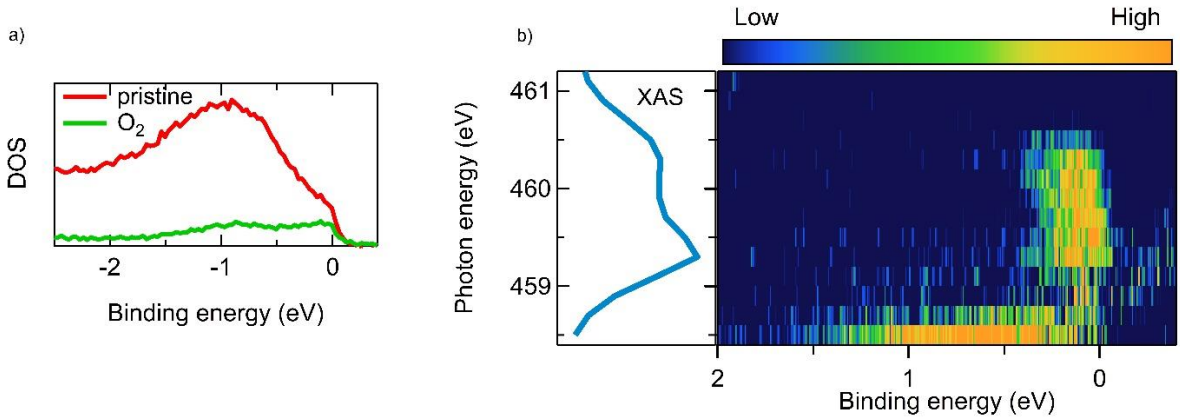

**Supplemental Figure S10** DOS measured of anatase films deposited on STO. The behaviour of IG and 2DEG upon oxygen dosing, as well as the resonating behaviour as function of the photon energy is consistent with the behavior observed for LAO, thus confirming that the results obtained are intrinsic features of the TiO$_2$ films. (**a**) Effect of the oxygen dosing on the photoemission intensity (momentum integrated between k$_x$ = -0.2 and 0.2 A$^{-1}$) of a 20nm-thick TiO$_2$/ Nb-SrTiO$_3$ sample acquired in the 2$^{nd}$ BZ using hv =459.7 eV circularly polarised radiation, at 90 K. (**b**) ResPES acquired around the 2BZ of the same TiO$_2$/ Nb-SrTiO$_3$ sample upon oxygen dosing (P =

$1 \times 10^{-9}$ mbar) to better decouple the IGs and the 2DEG signals. Circularly polarized light was varied across the Ti $L_3$-eg resonance (total electron yield in left panel). Temperature was set to 90 K. The colour map (right panel) displays the momentum-integrated (between $k_x$ = -0.2 and 0.2 Å$^{-1}$) photoemission intensity as a function of the photon energy and of the binding energy of the electrons.

SUPPLEMENTARY REFERENCES

1. S. Moser, L. Moreschini, J. Jaćimović, O. S. Barišić, H. Berger, A. Magrez, Y. J. Chang, K. S. Kim, A. Bostwick, E. Rotenberg, L. Forró, M. Grioni, *Tunable Polaronic Conduction in Anatase TiO2*, Physical Review Letters **110** (19), 196403 (2013).
2. A. F. Santander-Syro, O. Copie, T. Kondo, F. Fortuna, S. Pailhès, R. Weht, X.G. Qiu, F. Bertran, A. Nicolaou, A. Taleb-Ibrahimi, P. Le Fèvre, G. Herranz, M. Bibes, N. Reyren, N.; Apertet, Y.; Lecoeur, P.; Barthélémy, A.; Rozenberg, M. J., *Two-dimensional electron gas with universal subbands at the surface of SrTiO3*, Nature **469**, 189 (2011).
3. P. D. C. King, S. McKeown Walker, A. Tamai, A. de la Torre, T. Eknapakul, P. Buaphet, S. K. Mo, W. Meevasana, M. S. Bahramy, F. Baumberger, *Quasiparticle dynamics and spin–orbital texture of the SrTiO3 two-dimensional electron gas*, Nature Communications **5**, 3414 (2014).
4. Z. Wang, Z. Zhong, S. McKeown Walker, Z. Ristic, J. Z. Ma, F. Y. Bruno, S. Riccò, G. Sangiovanni, G. Eres, N. C. Plumb, L. Patthey, M. Shi, J. Mesot, F. Baumberger, M. Radovic, *Atomically Precise Lateral Modulation of a Two-Dimensional Electron Liquid in Anatase TiO2 Thin Films,* Nano Letters **17** (4), 2561-2567 (2017).
5. T. C. Rödel, F. Fortuna, F. Bertran, M. Gabay, M. J. Rozenberg, A. F. Santander-Syro, P. Le Fèvre, *Engineering two-dimensional electron gases at the (001) and (101) surfaces of TiO2 anatase using light.* Physical Review B **92** (4), 041106 (2015).
6. Y. Aiura, K. Ozawa, E.F. Schwier, K. Shimada, K. Mase, *Competition between Itineracy and Localization of Electrons Doped into the Near-Surface Region of Anatase TiO2*, The Journal of Physical Chemistry C **122** (34), 19661-19669 (2018).
7. J. Gabel, M. Zapf, P. Scheiderer, P. Schütz, L. Dudy, M. Stübinger, C. Schlueter, T. L. Lee, M. Sing, R. Claessen, *Disentangling specific versus generic doping mechanisms in oxide heterointerfaces*, Physical Review B **95** (19), 195109 (2017).
8. N. C. Plumb, M. Salluzzo, E. Razzoli, M. Månsson, M. Falub, J. Krempasky, C. E. Matt, J. Chang, M. Schulte, J. Braun, H. Ebert, J. Minár, B. Delley, K. J. Zhou, T. Schmitt, M. Shi, J. Mesot, L. Patthey, M. Radović, *Mixed Dimensionality of Confined Conducting Electrons in the Surface Region of SrTiO3*, Physical Review Letters **113** (8), 086801 (2014).
9. S. M. Walker, F. Y. Bruno, Z. Wang, A. de la Torre, A. Riccó, A. Tamai, T. K. Kim, M. Hoesch, M. Shi, M. S. Bahramy, P. D. C. King, F. Baumberger, *Carrier-Density Control of the SrTiO3 (001) Surface 2D Electron Gas studied by ARPES,* Advanced Materials **27** (26), 3894-3899 (2015).